%% file: tri3d.tex
\documentclass[12pt]{iopart}

\usepackage{iopams}
\usepackage{graphicx}
\usepackage{epsfig}

\newcommand{\sgn}{\;\mbox{sgn}}

\newcommand{\eps}{\varepsilon}
\newcommand{\mx}{{\mbox{\tiny max}}}
\newcommand{\SP}{{\mbox{\tiny SP}}}

\newcommand{\Num}{{\mbox{\tiny Num}}}

\begin{document}

\bibliographystyle{unsrt}

\title[Spectral properties of right triangles]
{Generic spectral properties of right triangle billiards}

\author{Thomas Gorin}
\address{Centro de Ciencias Fisicas, UNAM, Cuernavaca, Morelos, Mexico}
\address{Centro Internacional de Ciencias, Cuernavaca, Morelos, Mexico}

\ead{gorin@cicc.unam.mx}

\begin{abstract}
This article presents a new method to calculate eigenvalues of right triangle 
billiards. Its efficiency is comparable to the boundary integral method and 
more recently developed variants. Its simplicity and explicitness however allow 
new insight into the statistical properties of the spectra. We analyse 
numerically the correlations in level sequences at high level numbers ($>10^5$) 
for several examples of right triangle billiards. We find that the strength of
the correlations is closely related to the genus of the invariant surface of 
the classical billiard flow. Surprisingly, the genus plays and important 
r\^ ole on the quantum level also. Based on this observation a mechanism 
is discussed, which may explain the particular quantum-classical correspondence 
in right triangle billiards. Though this class of systems is rather small, it 
contains examples for integrable, pseudo integrable, and non integrable 
(ergodic, mixing) dynamics, so that the results might be relevant in a more 
general context.
\end{abstract}

\submitto{\JPA}
\pacs{03.65.GE, 03.65.Sq, 05.45.-a}

 \maketitle

\section{\label{I} Introduction}

Polygon billiards have been studied both classically and quantum mechanically 
for roughly twenty years now \cite{RicBer81}. These systems are situated right 
on the borderline between integrability and chaos. They are usually
divided into two classes: the rational polygon billiards where all vertex 
angles are rational multiples of $\pi$, and the irrational ones where at least 
one vertex angle is an irrational multiple of $\pi$. 

In the first case, there exist two constants of motion, so that one
would expect integrability. However, due to singularities in the billiard flow,
the invariant surface of the flow is not necessarily a torus (with genus 
$g=1$), but may be of a more complicated topology ($1\le g<\infty$). This 
produces a very complicated classical dynamics (see: 
\cite{Gut86,Gut96,KenSmi00} and references therein). The systems are called 
integrable if $g=1$ and pseudo integrable \cite{RicBer81} otherwise. 

In the second case (the irrational polygon billiards), there is no second 
constant of motion. These systems are typically ergodic \cite{Gut86} and 
probably weakly mixing \cite{ArtCas97,CasPro99}, though the Kolmogorov-Sinai 
entropy \cite{LichLieb83} is always zero. \\
 
Quantum and semiclassical calculations have been performed from the very 
beginning \cite{RicBer81,Gau87,Shu93,Mil94}, but only recently 
\cite{Bog99,CasPro99b} it became possible to calculate sufficiently large 
level sequences at sufficiently high energies, such that correlation properties 
could be analysed directly. There are fundamental open questions:
\begin{itemize}
\item[(i)]{Do the correlations in the spectra of polygon billiards eventually
become stationary at sufficiently high energy?}
\item[(ii)]{Are there families of polygon billiards with common statistical 
properties (universality)?}
\item[(iii)]{What is the signature of classical pseudo integrability in the
quantum spectrum (quantum-classical correspondence)?}
\end{itemize}
On the one hand, there has been numerical evidence \cite{CasPro99b}, that
at very high energies the spectra of irrational triangle billiards are
statistically similar to spectra taken from the Gaussian Orthogonal Ensemble 
(GOE). On the other hand, based on the numerical study of the spectra of 
several rational right triangle billiards, it was proposed that pseudo 
integrability implies so called ``intermediate statistics'' \cite{Bog99}. For 
the nearest neighbour distribution \cite{Boh89} this means: linear increase at 
small spacings (as in the GOE case) and exponential fall-off at large spacings 
(as for a random Poissonian sequence). Intermediate statistics has also been 
found in the context of disordered systems at the metal-insulator transition 
point \cite{Shk93,BraMon98,VarBra00}, which might indicate some relationship 
between both classes of systems. \\

This paper is mainly concerned with question (iii). We consider the 
one-parameter family of right triangle billiards, labeled by the value of the
smallest vertex angle $0< \alpha \le \pi/4$. For this class, a secular equation 
is derived, which identifies the eigenvalues as zeros of the determinant of a 
particular matrix $K(E)$. Though the matrix is infinite, its elements are given 
explicitly by very simple expressions. This makes $K(E)$ an ideal point of 
departure for numerical and analytical studies.

The most obvious characteristic of rational polygon billiards is the genus $g$ 
of the invariant surface of the classical Hamiltonian flow (the irrational polygon 
billiards can be included, setting $g=\infty$). Hence we will investigate in
detail the relation between $g$ and the correlations in the quantum spectra.
In the numerical part, level sequences are calculated at absolute level numbers 
$>10^5$ for various examples of right triangle billiards. This provides valuable 
complementary information to recent results from Bogomolny et al. \cite{Bog99}.
In the analytical part, the matrix $K(E)$ itself is considered. Though $K(E)$ 
is a pure quantum mechanical object, it is shown that $g$ and $\gamma$ (which 
is closely related to $g$) play a crucial r\^ ole for iterated mappings of the 
form $\Psi(n)= K^n(E) \Psi(0)$. Based on this observation, a mechanism is 
proposed, which can explain the connection between the genus $g$ and the 
correlation properties of the quantum spectrum. \\

In section~\ref{SE} a secular equation is derived for the calculation of the 
eigenvalues of right triangle billiards. It is used in section~\ref{LS} to
obtain and analyse the level spacing distributions for several right triangles. 
In section~\ref{CM} we analyse the properties of the matrix $K(E)$ itself, and
we discuss the r\^ oles of the two classical parameters $g$ and $\gamma$ in 
this context. The conclusions are presented in section~\ref{C}.

\section{\label{SE} Secular equation}

Our point of departure is the observation, that any right triangle can be 
obtained from cutting an appropriate rectangle along its diagonal. This is 
used to derive a secular equation of drastically reduced dimension for the 
eigenvalues of the right triangle billiard.

Let $H_0$ be the Hamiltonian for the rectangle billiard with sides $a$ and $b$.
Fixing the length scale by: $a^2+b^2= \pi^2$, the angle 
$\alpha : \tan\alpha = b/a$ suffices to characterize the system completely. 
Choosing an arbitrary corner of the rectangle billiard as the origin of a 
Cartesian coordinate system, its eigenvalues and the corresponding eigenfunctions 
may be written as follows:
\begin{eqnarray}
\eps(n,m) &= \frac{1}{2}\left(\frac{n^2}{\cos^2\alpha} +
\frac{m^2}{\sin^2\alpha}\right) \; , \quad n,m \ge 1 \label{SE_epsrec}\\
\Phi_{nm}(x,y) &= \frac{2}{\sqrt{ab}} \sin\!\!\left(\frac{\pi}{a} nx\right) 
\sin\!\!\left(\frac{\pi}{b} my\right) \; .
\label{SE_Phirec}\end{eqnarray}
Consider the total Hamiltonian $H$:
\begin{equation}
H = H_0 + \eta \; W \; ,\quad 
W = \delta\!\!\left( \frac{x}{a}-\frac{y}{b}\right) \; ,
\label{SE_Htri}\end{equation}
where the potential $\eta W$ is used to cut the rectangle billiard into two
congruent right triangle billiards (a similar cut potential, though in a 
different context, has been used in \cite{Lew90}). As $\eta$ increases from $0$ 
to $\infty$, the spectrum of $H$ changes from the spectrum of the rectangle 
billiard (\ref{SE_epsrec}) to the doubly degenerated spectrum of the two 
triangle billiards. For any $\eta$, the Hamiltonian $H$ is invariant under 
point reflection, so that the matrix representation of $H$ in the eigenbasis 
of $H_0$ is block diagonal. One block is spanned by the odd basis states 
$\{ \Phi_{nm} |\; n+m : {\rm odd} \}$ and the other by the even ones 
$\{ \Phi_{nm} |\; n+m : {\rm even} \}$. Both blocks can be diagonalised
independently, leading to the same sequence of eigenvalues, which causes the
degeneracy mentioned above. \\

In what follows we will work in the odd basis only. Let $q= n+m$ and 
$p= n-m$, and order the states (\ref{SE_Phirec}) with increasing $q$, and for
equal $q$, with increasing $p$. Consider the subset of states with fixed $q$ 
and $p= -q+2,\ldots,q-2$ as one block. Then truncating the basis at a maximal 
$q$-value $q_\mx$, one obtains $M= (q_\mx -1)/2$ blocks with $q-1$ states in each 
block (note that $q$ and $p$ are odd). In total this gives $N= (q_\mx^2 -1)/4$ 
basis states. In this reordered basis, the matrix elements of $W$ are given by:
\begin{equation}
\eqalign{
W_{qp;q'p'} &= \int_0^a\rmd x\int_0^b\rmd y \; \Phi_{nm}(x,y) \Phi_{n'm'}(x,y) 
\; \delta\left(\frac{x}{a}-\frac{y}{b}\right) \\
&= \; \frac{1}{2} \{ \delta(|p|-|p'|) + \delta(q-|p'|) 
+ \delta(q'-|p|) + \delta(q-q')\} \; ,
}
\end{equation}
where $n= (q+p)/2,\; m=(q-p)/2$ (and similarly for the primed indices). For 
given $q_\mx$ the truncated matrix $W^{(N)}$ has only two distinct eigenvalues: 
$0$ and $M+1$, and the eigenspace of the latter has dimension $M$ (in other 
words: rank$[W^{(N)}] = M$). All eigenvectors with eigenvalue $M+1$ can be 
calculated explicitly, and after proper normalization we collect them (as 
column vectors) in the rectangular matrix $V$:
\begin{equation}
V_{k;qp} = \frac{1}{\sqrt{M+1}}
\left\{ \begin{array}{ll}
0 \quad &: n_q < k \\
\sqrt{\frac{k+1}{k}} \quad &: n_q = k \\
\frac{1}{k(k+1)} \quad &: n_q > k \; ,\; n_p < k+1 \\
-\sqrt{\frac{k}{k+1}} \quad &: n_q > k \; ,\; n_p = k+1 \\
0\quad  &: n_q > k \; ,\; n_p > k+1 \end{array}\right.  \; , 
\label{AVres}\end{equation}
where $k=1,\ldots,M$, $n_q= (q-1)/2$ and $n_p= (|p|+1)/2$. The truncated total 
Hamiltonian may now be written as follows:
\begin{equation}
H^{(N)}= H_0^{(N)} + \eta (M+1) \; VV^T \; .
\label{SE_Htru}\end{equation}
Dividing the Schr\"odinger equation $(E-H^{(N)})\Psi =0$ by $E-H_0^{(N)}$, one
arrives after a few algebraic manipulations at the desired secular equation. It
determines the eigenvalues of $H^{(N)}$ as the zeros of the following determinant:
\begin{equation}
0 = \det\left( 1 + \eta \tilde K^{(M)}(E) \right) \; , \qquad
\tilde K^{(M)}(E) = (M+1) \; V^T \frac{1}{E-H_0^{(N)}} V \; .
\label{SE_Sec}\end{equation}
Taking the limit $\eta\to\infty$, the unit matrix in the first equation of 
(\ref{SE_Sec}) can be neglected, and one gets:
\begin{equation}
\det \tilde K^{(M)}(E) = 0 \; .
\label{SE_Sec2}\end{equation}
The advantage of this equation, is the reduced dimension $M\ll N= M(M+1)$. 
Such a reduction is typical for a boundary integral method (see for example 
\cite{VerSar95}). The matrix elements of $\tilde K^{(M)}$ are given by the 
following expression:
\begin{equation}
\tilde K_{ij}^{(M)} = (M+1) \sum_{q=1}^{q_\mx} \sum_{p=-q+2}^{q-2} 
\frac{V_{i,qp}\; V_{j,qp}}{E- \eps\left(\frac{q+p}{2},\frac{q-p}{2}\right)} 
\; , \quad q,p:{\rm odd} \; .
\label{SE_tilK}\end{equation}
Being only interested in the zero eigenvalues of $\tilde K^{(M)}(E)$, any 
(symmetric) similarity transformation $K^{(M)} = L^T \tilde K^{(M)} L$ may be 
applied. The following choice for $L$ simplifies the problem considerably:
\begin{eqnarray}
L &=& {\rm diag}(1,\ldots,1/M) \, \left( \begin{array}{cccc}
1 & -1 \\
 & \ddots & \ddots \\
 & & \ddots & -1 \\
 & & &  1 \end{array}\right) \, 
{\rm diag}\!\!\left(1,\ldots,\sqrt{M(M+1)}\right) \, . \nonumber\\ 
\quad &\quad &\quad
\end{eqnarray}
The resulting matrix $K^{(M)}(E)$ is defined in the same way as $K(E)$ in the 
equations~(\ref{SE_Kfin1})-(\ref{SE_Kfin3}) below, but with the coefficients 
$D_j= \sum_{i=0}^M d_{ij}$. Only in the limit $M\to \infty$, the expression for
$D_j$ simplifies to the formula~(\ref{SE_Kfin4}), as can be shown using the 
partial fraction expansion of the {\it cot} function \cite{AbraSteg64}. \\

To summarize, the right triangle spectrum is calculated using the secular
equation ${\rm det}[K(E)]= 0$, where $K(E)$ is constructed as follows:
\begin{equation}
K(E) =  K_F(E) + K_D(E) \; .
\label{SE_Kfin1}\end{equation}
The matrix elements of $K_F$ are given by:
\begin{equation}
\eqalign{
[K_F]_{ij} = d_{ij} - d_{i,j+1} -d_{i-1,j} + d_{i-1,j+1}  \; , \\
d_{ij} = d_{ij}^+ + d_{ij}^- \; , \qquad
d_{ij}^\pm = \frac{1}{e - q^2 - p^2 \pm 2qp \cos 2\alpha}  \; ,
}
\label{SE_Kfin2}\end{equation}
where the scaled energy $e= E/(2\sin^2 2\alpha)$ is used, and $q= 2i+1$, and 
$p= 2j-1$. Note, that $q^2+p^2\mp 2qp\, \cos(2\alpha) =
2\sin^2(2\alpha)\;\eps[(q\pm p)/2,(q\mp p)/2]$, thus $q$ and $p$ may still be 
regarded as auxiliary quantum numbers for the rectangle billiard $H_0$. The 
matrix $K_D$ is tridiagonal:
\begin{equation}
[K_D]_{jj} = D_j + D_{j+1} \; , \qquad
[K_D]_{j,j+1} = [K_D]_{j+1,j} = - D_j \; ,
\label{SE_Kfin3}\end{equation}
with the coefficients $D_j$, given by
\begin{equation}
D_j = \frac{\pi \; \sin\pi \omega}{2\omega 
\left( \cos\pi \omega + \cos\pi p \cos 2\alpha\right)}\; , \qquad
\omega = \sin 2\alpha \; \sqrt{2e- p^2} \; .
\label{SE_Kfin4}\end{equation}
Even though $\omega$ becomes imaginary for large values of $p$, the affected
functions: {\it sin} and {\it cos} convert into {\it sinh} and {\it cosh}, 
and finally the coefficient $D_j$ remains real. Its asymptotic behaviour for
large $j$ is: $D_j \sim \pi/(2|\omega|)$. \\

This result is the basis for the analysis in section~\ref{LS}~and 
section~\ref{CM}. The infinite matrix $K(E)$, as defined in
(\ref{SE_Kfin1})-(\ref{SE_Kfin4}), completely determines the spectrum 
of any right triangle billiard as the set of zeros of its determinant. For 
numerical purposes $K(E)$ must be truncated (see below), but one may get 
important information also from an analysis of the infinite matrix $K(E)$ 
itself.

For numerical calculations (section~\ref{LS}), $K(E)$ is truncated, keeping
only those elements $K_{ij}(E)$ for which $i,j \le M$. For meaningful results, 
$M$ must be at least so large, that $p_\mx^2 > 2e,\; p_\mx= 2M-1$ [see the 
definition of $p$ above equation~(\ref{SE_Kfin3})]. Experience shows, that for 
accurate results (error less than $1\%$ of the mean level spacing), one should 
increase the size of the matrix further by approximately 10\%. The zeros of 
$\det[K(E)]$ are identified, calculating the smallest eigenvalue in magnitude 
as a function of $E$. Using a standard root bracketing algorithm 
\cite{NumRec92} we find those points at which the smallest eigenvalue of 
$K(E)$ passes the zero axis. The eigenvalues of $K(E)$ are strictly decreasing 
functions of $E$, and this facilitates the root finding considerably. It allows
to take rather large steps (of the order of the mean level distance), without
running the risk to loose any roots.

\section{\label{LS} Level spacing distributions}

In the case of polygon billiards, the genus $g$ of the invariant surface of the
Hamiltonian flow is the most obvious parameter to characterize the classical 
dynamics \cite{Gut86}. Hence one may expect an influence of $g$ on the level 
statistics of the corresponding quantum system. In this section we investigate 
numerically whether the level statistics show a systematic dependence on $g$. 
For several rational and one irrational right triangle, we calculate sequences 
of $10^4$ levels starting at the absolute level number $10^5$ (Weyl's 
law is used to determine the corresponding energy). Note that even in this 
energy region the level statistics are usually not stationary. This has been 
demonstrated in \cite{CasPro99b} for several examples of rational and 
irrational triangle billiards. This should be kept in mind in the discussion
of the numerical results.

In the case of right triangle billiards, there is another relevant parameter 
intimately related to $g$ (see~\ref{AG}). This is $\gamma$, the smallest 
integer such that $2\alpha\,\gamma/\pi \in \mathbb{N}$ (in the irrational case, 
we set $\gamma = \infty$). It is shown in \ref{AG}, that $\gamma$ is the 
smallest number of rhombuses which must be glued together to form the 
invariant surface of the billiard flow. Moreover we find, that 
$g= {\rm int}(\gamma/2)$. Hence $\gamma$ implies a finer classification of 
the right triangle billiards than $g$ does. \\

\begin{table}
\caption{All rational right triangle billiards with $g\le 7$, referenced by 
their smallest vertex angle $\alpha/\pi = p/q$, and ordered with respect to 
$g$ and $\gamma$. The first two entries for $g= g_a= 1$ are the only integrable 
cases. The shaded entries refer to those cases analysed in \cite{Bog99}, and 
the gray-scale corresponds to the value of $g_a$ (introduced there) as 
indicated in the last column. \\ }
\begin{indented}
\item[] \includegraphics[scale=0.55]{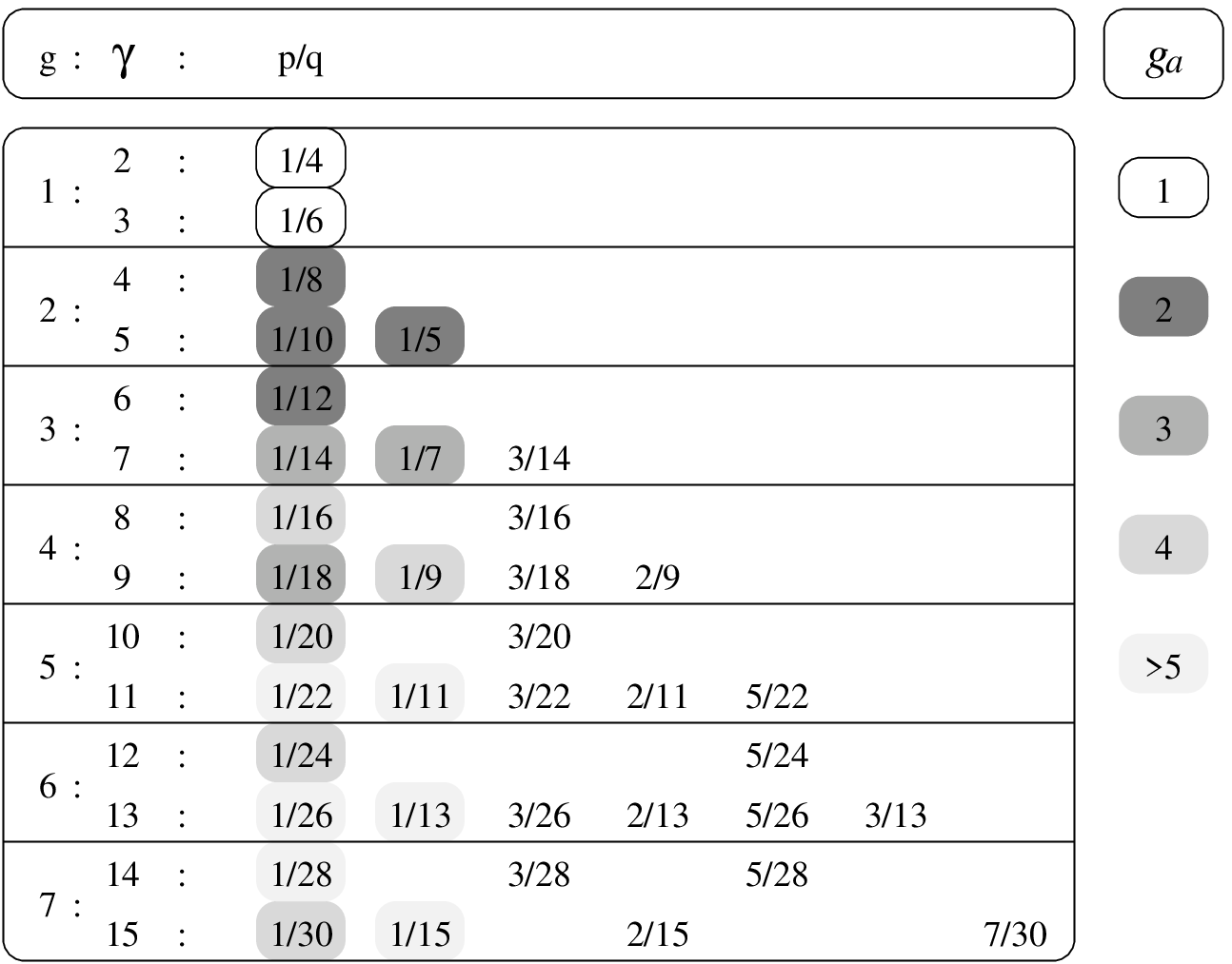} 
\end{indented}
\label{LS_gen2}\end{table}

Up to an irrelevant energy scale, all right triangles may be labeled and
identified through their smallest vertex angle $0< \alpha \le \pi/4$. For
rational right triangles one may also use the pair of relatively prime integers
$p/q= \alpha/\pi$. This is done in table~\ref{LS_gen2}, where all rational 
right triangles with $g\le 7$ are arranged with increasing $\gamma$ in the 
vertical direction, and with increasing $\alpha$ in the horizontal direction. 
The parameter $g_a$ is taken from \cite{Bog99}, where it is introduced as 
``arithmetical genus''. The entries under-laid with a gray shade have been 
analysed there. The gray-scale corresponds to different values of $g_a$, as 
indicated in the last column. Further below, we will compare our results with 
those of \cite{Bog99}. \\

\begin{figure}
\begin{center}
\input{deli12b.pstex_t}
\end{center}
\caption{Difference $\Delta I\, (s)$ of the integrated level spacing 
distribution to the semi-Poisson case. $\Delta I\, (s)$ is plotted for various
values of $\alpha$ as indicated in (a). The abbreviation ``irr'' refers to 
$\alpha= \pi(3-\sqrt{5})/4$. Whereas (b) shows the raw data, (a) shows the 
corresponding smoothed curves (details in the text). The GOE expectation 
($N\to\infty$ limit) is plotted in (a) and (b) as a dotted line.}
\label{LS_f1}\end{figure}
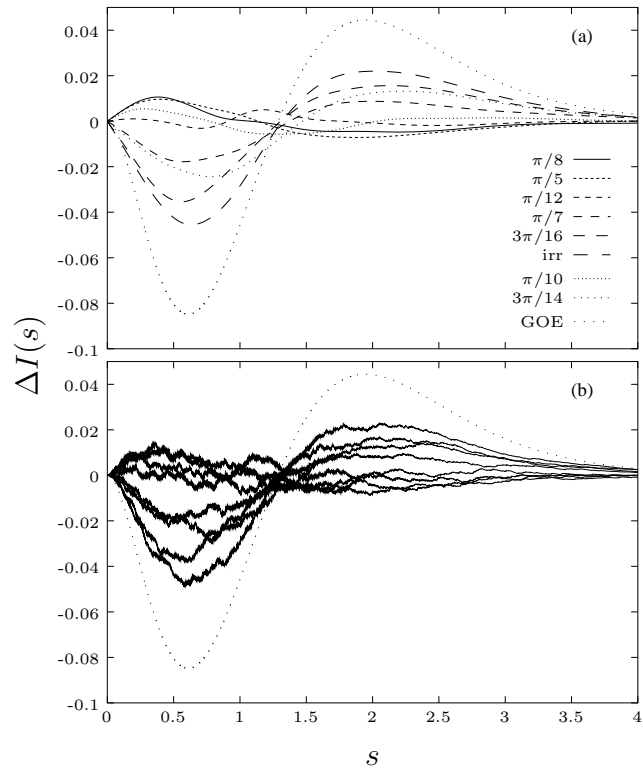

We analyse the level statistics by means of the nearest neighbour spacing 
distribution $P(s)$ (the spacings are normalized to unit mean). Recently the
question has been raised, whether the rational right triangles show 
intermediate statistics ({\it i.e.} a linear increase at small $s$ and 
exponential fall-off at large $s$). A simple analytical example is the so 
called ``semi-Poisson'' distribution \cite{Bog99}:
\begin{equation}
P_\SP(s) = 4s \; \rme^{-2s} \; .
\label{I_semiP}\end{equation}
Here, $P_\SP(s)$ is simply used as a conveniant reference to compare with.
The following quantity is plotted in figure~\ref{LS_f1}:
\begin{equation}
\Delta I\, (s) = \int_0^s\rmd s'\; \left\{ P_\Num(s') - P_\SP(s')\right\} \; .
\label{LS_DIs}\end{equation}
The theoretical curves in the figures~\ref{LS_f1}(a) and (b) show the result 
for an infinite GOE spectrum, where $P_\Num(s')$ is replaced by the 
corresponding level spacing distribution (taken from \cite{Haake91}). While 
figure~\ref{LS_f1}(b) shows the raw numerical data curves for various right 
triangle billiards, figure~\ref{LS_f1}(a) shows the corresponding smoothed 
curves, in order to allow the identification of all the cases shown. For the 
smoothing, ``natural smoothing splines'' have been used, as provided in 
\cite{gnuplot99}. 

\begin{figure}
\begin{center}
\input{CM.pstex_t}
\end{center}
\caption{Difference of the integrated level spacing distribution to the
semi-Poisson case. In (a) $\Delta I\, (s)$ is plotted for the rational right
triangle billiard with $\alpha/\pi= 1/8$, in (b) for the irrational one with 
$\alpha/\pi = (3-\sqrt{5})/4$. The solid lines show the result for
neighbored eigenvalues of the matrix $K(E)$, the dotted lines  show the result
for the triangle spectrum. The dashed lines show the theoretical curves
for the Poisson case (long dashed lines) and for the GOE case (short dashed 
lines).}
\label{CM_f3}\end{figure}
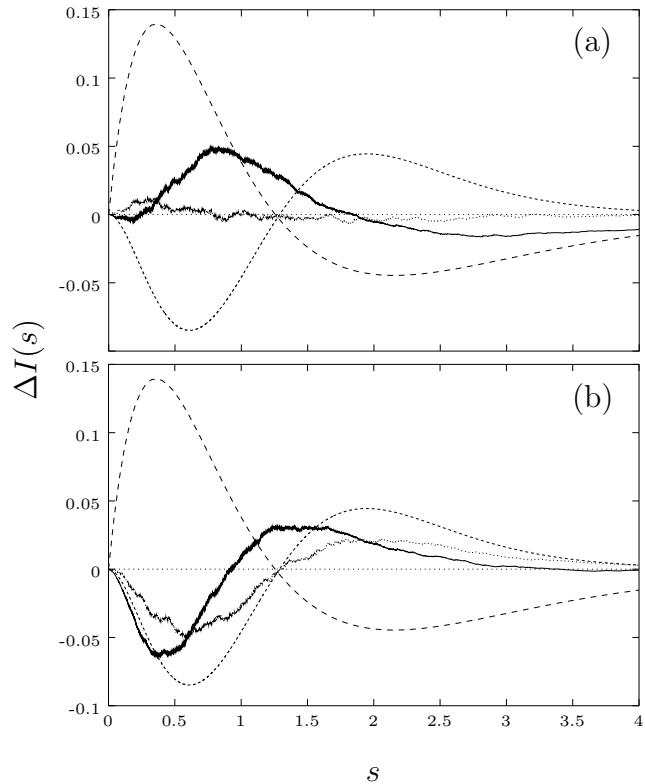

Let us first focus on the cases: $p/q= 1/8,1/5,1/12,1/7,3/16$ (which 
correspond to a successive increase of $\gamma$ from $4$ to $8$), and 
$\alpha/\pi = (3-\sqrt{5})/4$ (where $\gamma=\infty$). The $\Delta I$-curves 
for these cases are plotted in figure~\ref{LS_f1}(a) with a solid line and 
dashed lines of different dash lengths. Together with the results for the GOE 
and the Poisson ensemble (uncorrelated random sequence), they roughly span a 
one-parameter family of curves $\Delta I_\sigma(s)$. The parameter $\sigma$ may 
be called ``correlation strength'' and it may be calibrated, requiring that 
$\sigma=0$ gives the Poisson result (its graph is plotted in 
figure~\ref{CM_f3}), $\sigma=1$ the GOE result, and $\sigma=1/2$ the 
semi-Poisson result. Note, that $\Delta I_\sigma(s)$ is introduced solely to 
facilitate the discussion of our results, so that it is not necessary to be 
more specific.

The $\Delta I$-curve for the irrational right triangle billiard comes closest
to the GOE result. However, it remains almost in the middle between the 
semi-Poisson case and the GOE case ($\sigma \gtrsim 3/4$). Then follow the
cases $p/q= 3/16, 1/7$, and $1/12$, for which $\sigma$ decreases in 
approximately equal steps. The $\Delta I$-curve for $p/q=1/12$ is closest to 
the semi-Poisson result ($\sigma\approx 1/2$). The last two curves with 
$p/q= 1/5$, and $1/8$ tend slightly towards the Poisson result. They are so 
close to each other, that we would assign the same correlation strength to 
both of them ($\sigma\lesssim 1/2$). In all we may say, that the correlation 
strength $\sigma$ increases with increasing $\gamma$.

Finally we included two more cases: $p/q= 1/10$ and $3/14$. In 
figure~\ref{LS_f1}(a) the respective $\Delta I$-curves are plotted with dotted 
lines. Thus we can compare the $\Delta I$-curves for the $1/5$- and 
the $1/10$-triangle ($\gamma=5$), and the $\Delta I$-curves for the $1/7$- 
and the $3/14$-triangle ($\gamma=7$). Both cases show, that even for right
triangles with the same value for $\gamma$, the respective $\Delta I$-curves 
may differ considerably. The relation between $\gamma$ and the correlation 
strength is apparently not very strict (at least not in the energy range 
considered).

In order to check, that our conclusions do not depend on the particular choice
of the correlation measure, we repeated the analysis above, using the number 
variance $\Sigma^2(l)$ \cite{Boh89} instead of $\Delta I(s)$. The results 
were perfectly compatible, so that a more detailed discussion is omitted. \\

In the numerical analysis presented here, we concentrate on right triangle 
billiards with small values for $g$ and $\gamma$. The main reason is, that 
there is certainly an energy scale below which the quantum system cannot 
possibly ``recognize'', whether the two hypotenuse angles $\alpha$ and $\beta$
are rational or not. Without any knowledge about this scale, small $g$ 
triangles are probably the more reliable examples, for the study of quantum 
signatures of pseudo integrability. Hence the triangles for which we can 
compare our results with those of \cite{Bog99} are only a few: 
$p/q= 1/8,1/5,1/12$, and $1/7$.

The results obtained in \cite{Bog99} agree with those presented here, only up 
to a certain qualitative level. Beyond, we find that the correlation strength 
has decreased considerably in almost all cases. This may be due to the higher 
energy region considered here, which results in the rationality of the 
hypotenuse angles being more important. However, the $\Delta I$-curves of the 
first group of right triangles (with $4\le\gamma\le 6$) changed much less then 
the others, such that the separation between both groups has decreased. It 
seems that this separation was decisive for the introduction of the 
arithmetical genus $g_a$. The fact that this separation has become much 
smaller now, indicates that $g_a$ is probably not an appropriate alternative 
for $g$. \\

According to the numerical results presented in this section, it is possible 
to order the right triangle billiards with respect to the strength of the
correlations found in their spectra, which coincides with that of increasing 
$\gamma$. This finding confirms the general conjecture, that the genus of the 
invariant surface of the classical billiard flow determines the strength of 
the spectral correlations on the quantum level. Though the spectral 
correlations are apparently not stationary at currently accessible energies, 
the ordering seems to be energy independent, as long as the level sequences to
be compared, start with the same absolute level number.

\section{\label{CM} The elliptic map}

In the first part of this section it is shown, that the parameters $g$ and 
$\gamma$ (see \ref{AG}) associated with the classical dynamics of right 
triangle billiards, are important characteristics of the matrix $K(E)$ itself. 
On the one hand this may be surprising, because $K(E)$ arose from a pure 
quantum mechanical approach (see section~\ref{SE}), but on the other hand it 
is a strong indication for the importance of classical pseudo integrability on 
the quantum level. In the second part, we present a tentative explanation for 
the dependence of the spectral statistics on $g$ and $\gamma$. \\

\begin{figure}
\begin{center}
\includegraphics[scale=0.75]{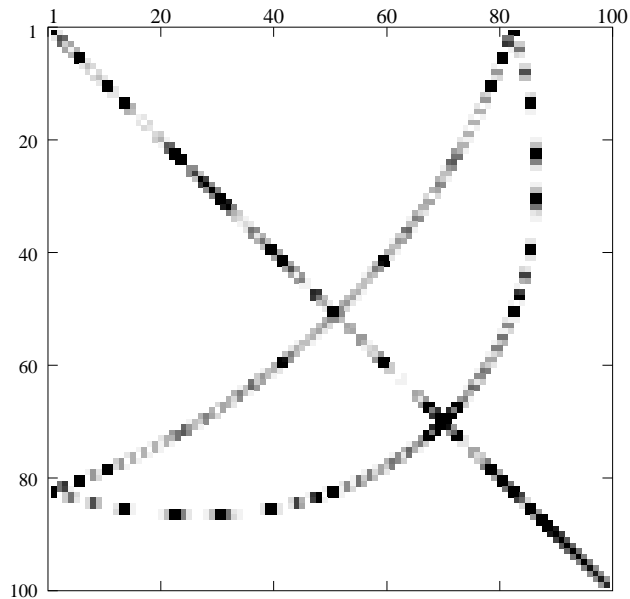}
\end{center}
\caption{Portrait of the matrix $K(E)$ for $\alpha = \pi/5$, and
$E= 1.26\times 10^4$. The gray-scale corresponds to the absolute value of the 
matrix elements.}
\label{CM_f1}\end{figure}

In figure~\ref{CM_f1} the matrix $K(E)$ is portrayed for a typical case. The 
gray-scale corresponds to the absolute value of the matrix elements. It can be 
seen, that most of the matrix elements have vanishingly small absolute values. 
Large absolute values can be found only along the diagonal and the first
off-diagonals, which are due to $K_D$, and on a ``moon''-like structure due to
$K_F$ [see equations~(\ref{SE_Kfin2})-(\ref{SE_Kfin4})]. The matrix elements 
$[K_F]_{ij}$ become large, whenever the pair of integers $(i,j)$ is close to 
the zero-line of one of the two functions
\begin{equation}
f_\pm(x,y)= e - 4\left( x^2 + y^2 \pm 2xy \cos 2\alpha \right) \; ,
\end{equation}
where $x,y$ are real, and positive, and $e$ is the scaled energy as used in
equation~(\ref{SE_Kfin2}). 

\begin{figure}[h]
\begin{center}
\input{kmatmap.pstex_t}
\end{center}
\caption{Schematic representation of the mapping $\vec y_1 = K(E) \vec y_0$
(details see text).}
\label{CM_fmap}\end{figure}
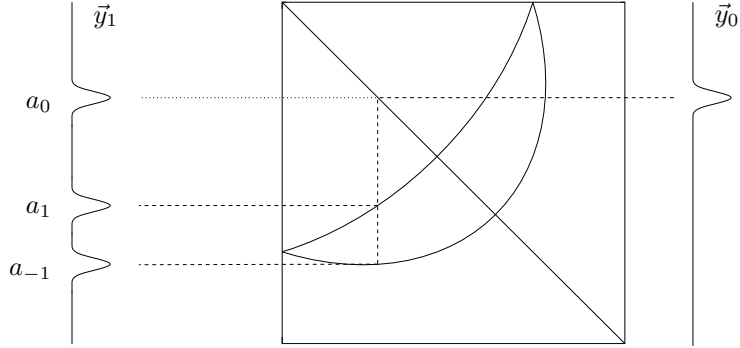

The action of $K(E)$ on a localized state may be described schematically by a 
double valued, symmetric map as shown in figure~\ref{CM_fmap}. The square in 
the middle represents the matrix $K(E)$ (cf. figure~\ref{CM_f1}). An initial 
state $\vec y_0$ localized at a given value $a_0$ is mapped to 
$\vec y_1 = K(E) \vec y_0$ localized at $\{ a_0, a_{-1}, a_1\}$, where $a_{-1}$ 
and $a_1$ are the two solutions of the equation $f_\pm(a_0,x)=0$, for $x>0$. 
Hence, the map $M$ associated with $K(E)$ may be defined as follows: 
$\{ a_{-1}, a_1\} = M a_0$. Let us call it the ``elliptic map''. Due to 
$f(x,y)=f(y,x)$, $a_0 \in M a_{-1}$ and $a_0 \in M a_1$. Consequently, an orbit 
of such a map may be viewed as a doubly connected chain:
\begin{equation}
\cdots\!\!\!\parbox{7.4cm}{\input{mapflow.pstex_t}}\; \cdots
\label{CM_mapflow}\end{equation}
According to that picture, the $n$-fold image $\vec y_n = K(E)^n \; \vec y_0$ 
has localization peaks at the positions $\{a_{-n},\ldots,a_0,\ldots,a_n\}$. 
Surprisingly, $M$ is isomorphic to the following, extremely simple map:
\begin{equation}
\varphi_{n\pm 1} = \varphi_n \pm 2\alpha \; ,
\label{CM_elli2}\end{equation}
where the result should be taken modulo $\pi$, such that it remains in the
interval $\left[ \right. -\pi/2,\pi/2\left. \right)$. This can be seen, using 
the following parametrisation of the curve 
$f_\pm(x,y) = 0$:
\begin{equation}
{x\choose y} = \sqrt{E}
{\sgn\left(\frac{\pi}{2}-2\alpha + \varphi\right) \; \cos (\varphi-2\alpha) 
\choose \cos (\varphi)} \; , \quad
\varphi\in \left[\right. -\pi/2,\pi/2\left. \right) \; .
\label{CM_elli}\end{equation}
Replacing $y$ by an arbitrary point $a_n$ of the map $M$, one gets the 
corresponding pair conjugated angles: $\cos(\pm\varphi_n)=a_n$. Replacing $x$ 
by $a_n$ one finds that $\pm\varphi_n$ must be mapped to 
$\pm\varphi_n -2\alpha$ (mod $\pi$). It follows, that
\begin{eqnarray}
  a_n &=& \cos(\pm\varphi_n) \nonumber\\
M a_n &=& \{a_{n-1},a_{n+1}\} =
\{\cos[\pm(\varphi_n-2\alpha)],\cos[\pm(\varphi_n+2\alpha)]\} \; .
\end{eqnarray}
From equation~(\ref{CM_elli2}) it follows that any orbit is restricted to a set 
of $\gamma$ points, where $\gamma$ is the smallest integer such that 
$2\alpha\,\gamma/\pi \in\mathbb{N}$. It is the same $\gamma$, which is 
introduced in \ref{AG} as the number of rhombuses forming the invariant 
surface of the billiard flow. Furthermore, the periodicity of the map $M$ is 
${\rm int}(\gamma/2)$ which is just the genus of that invariant surface. \\

Here in the second part of this section we discuss a mechanism which can 
explain the correspondence between the correlation properties of the quantum 
spectrum and the classical parameter $\gamma$. The line of reasoning is as 
follows:
\begin{itemize}
\item[1]{The correlation properties of the triangle spectrum at a given 
energy $E$ are closely related to the correlation properties of the eigenvalues 
of $K(E)$ in the vicinity of zero.}
\item[2]{At sufficiently high energy, $K_D$ (\ref{SE_Kfin1}) can be considered 
as a random tridiagonal matrix with eigenstates which are typically localized.}
\item[3]{The matrix $K_F$ (\ref{SE_Kfin1}) has such a form, that
repeated multiplications of an initially localized state with $K(E)$, produce 
an increasing number of copies at different positions. The positions are given 
by the elliptic map $M$.}
\item[4]{If $\alpha$ is rational, all orbits of the elliptic map are
periodic with period $g$ and restricted to $\gamma$ points. This leads to an 
approximate foliation of the Hilbert space into weakly coupled subspaces. 
For any irrational $\alpha$, the elliptic map is ergodic, and all basis states 
of the matrix $K(E)$ are strongly coupled.}
\end{itemize}
Point 3 has been treated in the first part of this section. The remaining
statements are discussed below.

\subsection*{Correlation properties of the eigenvalues of the matrix $K(E)$}

According to the secular equation derived in section~\ref{SE}, the triangle 
eigenvalues are given by those energies, at which at least one eigenvalue of 
$K(E)$ becomes zero. Therefore it seems plausible, that the correlation 
properties of the eigenvalues of $K(E)$ close to zero and the triangle 
eigenvalues are related. To verify this we calculate the spacing distribution 
for those two neighbouring eigenvalues which have opposite signs (without 
unfolding). The distribution is obtained from $10^4$ spacings, taken at 
equi-distant energies, with the step size adjusted to the mean level spacing of 
the corresponding triangle spectrum.

In figure~\ref{CM_f3} we compare the results. In the rational case (a) as 
well as in the irrational case (b), the $\Delta I(s)$-curves for the 
eigenvalue pairs of $K(E)$ and for the triangle spectra differ remarkably, 
though in (b) the agreement is somewhat better. However, at least 
qualitatively, the results are as expected: In the rational case, 
figure~\ref{CM_f3}(a), the eigenvalue statistics for $K(E)$ show indeed very 
weak correlations, much weaker even than the corresponding triangle spectrum. 
This can be seen from the $\Delta I(s)$-curve which clearly tends towards the 
Poisson result (note also the behaviour at large $s$). In the irrational case, 
figure~\ref{CM_f3}(b), both curves show relatively strong correlations.

\begin{figure}
\begin{center}
\input{f14fin.pstex_t}
\end{center}
\caption{The eigenstates of $K_D(E)$ with eigenvalues close to zero for a
typical case. On the left, the absolute value of the eigenvector coefficients,
plotted as a function of $i$, the index for the basis of $K(E)$. On the 
right, the corresponding eigenvalues plotted in a bar graph. 
$\alpha= \pi/5, E = 1.5\times 10^5$.}
\label{LS_f6}\end{figure}
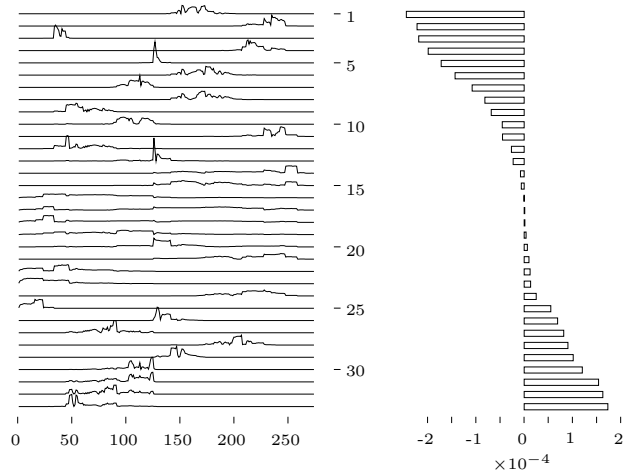

\subsection*{Localization of the eigenstates of $K_D$}

The matrix $K_D$ is constructed in a simple manner from the coefficients $D_j$
[see equations~(\ref{SE_Kfin3}) and (\ref{SE_Kfin4})]. Those oscillate as 
functions of $j$, the index of the basis of $K(E)$, more and more rapidly
while $p^2$ approaches $2e$. From a statistical point of view, it then seems
permissible to replace the arguments of the functions: {\it sin} and {\it cos}
by appropriate random variables. Even though the statistical properties of 
the matrix elements are very complicated, one may expect Anderson localization 
\cite{And58}.

In figure~\ref{LS_f6} we show for a typical case, a series of eigenstates of 
$K_D(E)$ ordered by their respective eigenvalues. Only those states with
eigenvalues in a small interval around zero are shown. Many eigenstates are 
apparently localized. However, others are not, and spread over a wide range of
basis states. Usually those fluctuate only weakly and slowly and their
eigenvalues decrease very slowly with energy (not shown). Their r\^ ole is 
still unclear, and will be the subject of future studies.

\subsection*{Foliation of the Hilbert space}

Acting repeatedly with $K(E)$ on an initially localized state $\vec y_0$, the 
first $g$ images will spread and localize at points 
$\{a_0,\ldots,a_{\gamma-1}\}$ (here $a_{\gamma-1}$ is identical to $a_{-1}$), 
as discussed in the first part of this section. Then, due to the periodicity of 
the elliptic map, subsequent images spread only slowly away from these points. 
In the ideal case the spreading would stop due to Anderson localization, 
giving rise to an invariant subspace. In the same way, an initial state 
localized in a different part of the Hilbert space, would lead to another 
invariant subspace, and so on -- until possibly the whole Hilbert space would 
have been decomposed into invariant subspaces. In the real system, such a 
foliation of the Hilbert space occurs only approximately, and the subspaces 
become weakly coupled. Nevertheless one may expect, that
correlations are to some extent suppressed due to this mechanism.

For irrational $\alpha$, where the elliptic map is ergodic, an initially
localized state will spread out [by repeated multiplication with $K(E)$]
into the whole Hilbert space. No foliation of the Hilbert space can occur, and 
one should expect correlations of similar strength as in the GOE case. \\

\section{\label{C}Conclusions}

We derived a new kind of secular equation for the determination of the spectra
of right triangle billiards. It involves the diagonalisation of the matrix 
$K(E)$ which has a particularly simple and transparent structure. Based on 
this equation we calculated spectra at level numbers $>10^5$ for various 
examples of right triangle billiards, which shows the efficiency of the new
method.

We found a clear correspondence between 
the genus $g$ (or the related parameter $\gamma$) of the invariant surface of 
the classical billiard flow and the strength of the correlations in the quantum 
spectrum. While for small $g$ the spectral statistics is close to semi-Poisson
(with a slight tendency towards Poisson), it approaches the GOE 
statistics when $g$ is increased. Our numerical results together with similar 
studies \cite{Bog99,CasPro99b} suggest that the spectral correlations are not 
stationary at currently accessible energies, but that the ordering with 
increasing correlation strength and its correspondence to $g$ is conserved.

In the second part of the paper, we found that the classical parameters $g$ 
and $\gamma$ are characteristic quantities for the matrix $K(E)$ itself. For
rational right triangle billiards, where $g$ is finite, one gets an approximate 
foliation of the Hilbert space into invariant subspaces. The size of the
subspaces scales with $\gamma$. Based on this observation, we discussed a 
mechanism which can explain the influence of $g$ and $\gamma$ on the level 
statistics of right triangle billiards. \\

The definition of $\gamma$ in \ref{AG} can be generalized to arbitrary polygon
billiards as follows: $\gamma$ is the smallest number of identical polygons
which must be glued together to form the invariant surface of the billiard
flow. In this general case, $g$ and $\gamma$ must possibly be considered as
independent parameters. Which of them is then more relevant for the
correlations in the quantum spectrum? Considering the r\^ oles of $g$ and 
$\gamma$ in the description of the matrix $K(E)$, it seems that this is
$\gamma$ (the size of the approximate invariant subspaces of $K(E)$ 
scales with $\gamma$).
%

\ack
The author thanks T.~Prosen for providing numerical triangle spectra, 
G.~Casati and T.~Prosen for making available their work prior to publication, 
and F.~Leyvraz and T.~Prosen for valuable discussions.

\appendix

\section{ \label{AG} The invariant surface for the classical billiard flow}

There is an elegant way to represent a trajectory moving in a polygon 
billiard, which is particularly useful to construct the invariant surface of
the classical billiard flow. It consists in drawing the trajectory as a 
straight line, and reflecting the billiard (instead of the trajectory) each 
time the boundary is hit \cite{Gut86}. In the case of rational polygons all
possible trajectories can produce only a finite number of differently oriented
copies of the original polygon. Then there is a general recipe of how to glue 
these copies together, in order to obtain the invariant surface. 

\begin{figure}
\begin{center}
\includegraphics[scale=0.55]{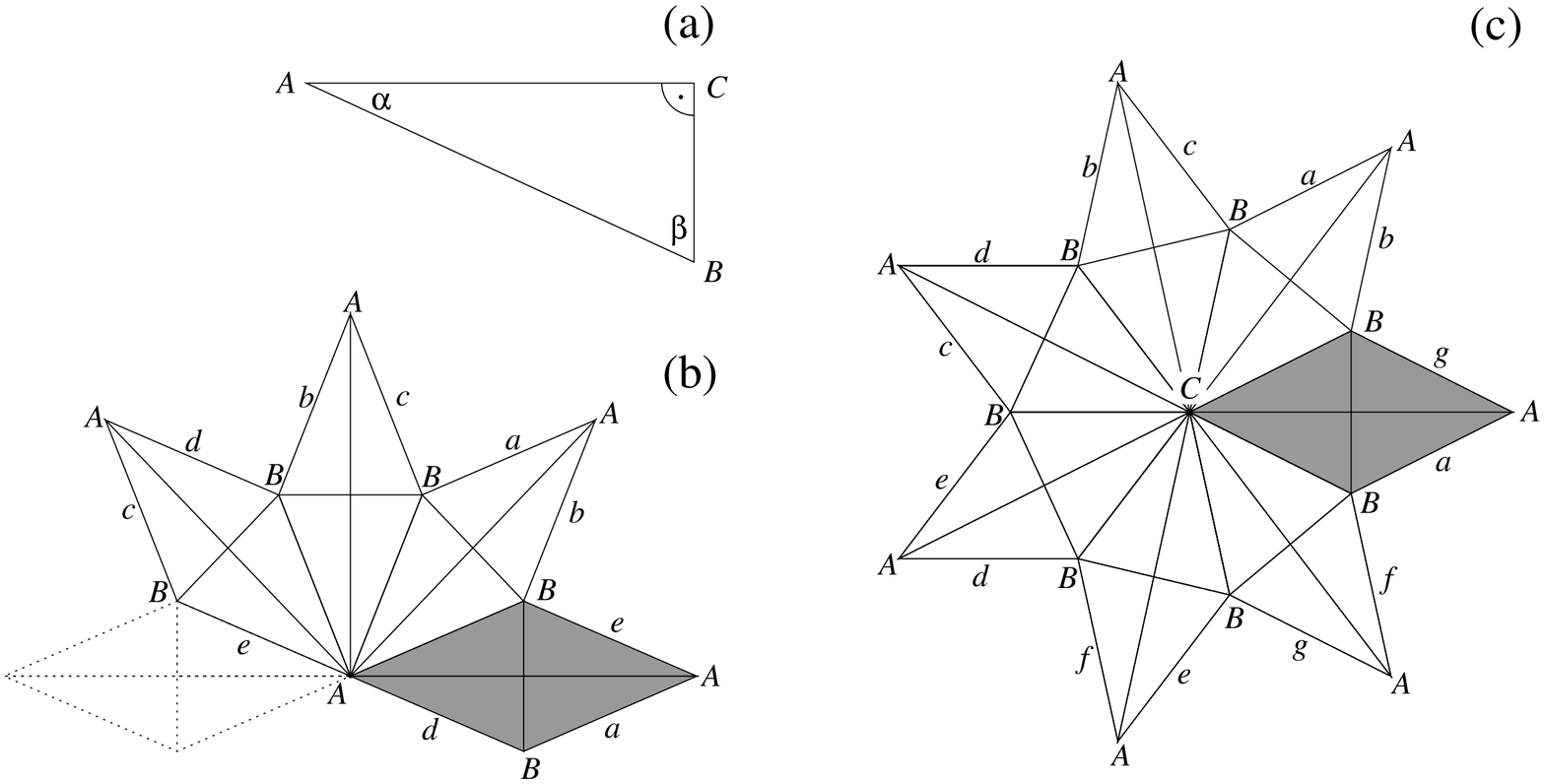}
\end{center}
\caption{(a) Typical right triangle. (b) Invariant surface of the billiard flow
for the $1/8$-triangle. The dotted rhombus does not form part of the invariant
surface. (c) Invariant surface of the billiard flow for the $1/7$-triangle.
In (b) and (c): the initial rhombuses, with which the rosette constructions 
are started are filled. Vertices are labeled by capital letters, edges by 
lower case letters. However, only such edges or vertices are labeled, which 
must be identified with one another, to obtain the invariant surface.}
\label{AP_fr}\end{figure}

For rational right triangles one may follow a more explicit construction 
scheme, which leads to a particularly simple invariant surface; the 
``rosette''. It is constructed as follows: Start with a right triangle as 
depicted in figure~\ref{AP_fr}(a). Reflecting the triangle on the side $AC$, 
the image on the side $BC$ and that image again on the side $AC$, produces a 
rhombus [see figure~\ref{AP_fr}(b) and (c)]. This rhombus is rotated around 
point $A$ by the angle $2\alpha$ (counter-clockwise) at each step. Stop one 
step before arriving at the original rhombus or its point reflected image [see 
figure~\ref{AP_fr}(b)]. Note, that the resulting surface may wind several times 
around $A$. As it is shown below, the resulting surface can be closed by
identifying open edges with one another, which gives the invariant surface.

The number of rotations by $2\alpha$ is just the smallest integer $\gamma$ 
such that $2\alpha\, \gamma/\pi\in\mathbb{N}$. We could equally well rotate
the rhombus step-wise by $2\beta$ around point $B$ [see figure~\ref{AP_fr}(b) 
and (c)], resulting in a different representation of the invariant surface.
However, as shown below, the number of rotations (or rhombuses) necessary to 
close the invariant surface is the same.

Proof: Let $p/q= \alpha/\pi$ and $p'/q'= \beta/\pi$ with $p,q$ and $p',q'$
relatively prime.
\begin{equation}
\eqalign{
2\gamma\; \frac{p}{q} = l = 
2\gamma \left( \frac{1}{2} - \frac{p'}{q'}\right) \; &\Rightarrow\; 
2\gamma\; \frac{p'}{q'} = \gamma - l \; , \\
2\gamma'\; \frac{p'}{q'} = l' = 
2\gamma' \left( \frac{1}{2} - \frac{p}{q}\right) \; &\Rightarrow\; 
2\gamma'\; \frac{p}{q} = \gamma' - l' \; .
}
\label{Aproof}\end{equation}
Consider the first line of (\ref{Aproof}). As $2 p < q$ it follows
$\gamma - l > 0$. But $\gamma'$ is the smallest integer such that 
$2\gamma'\; p'/q' \in\mathbb{N}$, hence: $\gamma' \le \gamma$. The same
argument applied to the second line of (\ref{Aproof}) shows: 
$\gamma \le \gamma'$. Therefore: $\gamma = \gamma'$. \\

It remains, to prove that the procedure above gives indeed a representation
of the invariant surface, and to calculate its genus $g$. To this end we show 
that all free edges of the rosette can be identified with one another. Then the rhombuses define a triangularisation of the invariant surface, and counting all 
faces $F$, edges $E$ and vertices $V$ of the triangularisation, we obtain the 
genus via the Euler characteristic \cite{Gut86}: $g= 1-\chi/2, \chi= V-E+F$. 

The edges may be divided into inner edges, which are connected to the center of 
the rosette, and outer edges, which are not connected to the center. Let us 
label both groups counter-clockwise by e$_1$,\ldots and e$_1'$,\ldots 
respectively, beginning with the lower edges of the initial rhombus (see 
figure~\ref{AP_fr}, but note that the labels shown there are different, and 
used only to identify different edges in the representation of the invariant
surface).

Let us first discuss the case, where $q$ is odd. Then $\gamma = q$ and the 
rosette winds $p$ times around its center before the last inner edge can be
identified with the first one [see figure~\ref{AP_fr}(c)]. In order to
identify all outer edges e$_1$,\ldots,e$_{2\gamma}$ pairwise with one 
another, observe that for $j$ :odd, a trajectory leaving the surface crossing 
e$_{j+3}$ would enter a triangle which is the parallel translation of the 
triangle with the hypotenuse e$_j$. By consequence, both edges can be 
identified. Hence one may identify the following edges: e$_1 \equiv$ e$_4$, 
e$_3 \equiv$ e$_6$, \ldots, e$_{2\gamma-3} \equiv$ e$_{2\gamma}$, and there are 
only two open outer edges left: e$_2$ and e$_{2\gamma-1}$ which can be 
identified with one another on the same grounds. 

The vertices in the representation of the invariant surface have to be 
identified taking into account that edges identified previously have the same 
initial and end points (the triangle connected to the edge defines an 
orientation). In this manner it is shown that for $q$ :odd, the rosette 
(invariant surface) has $\gamma$ faces, $2\gamma$ edges and $3$ vertices. 
Hence $\chi= 3-\gamma$ and $g= (\gamma-1)/2$.

If $q$ is even, then $p$ (relatively prime) is odd and the rosette winds $p/2$
times around its center [see figure~\ref{AP_fr}(b)] which means, that we have
also two open inner edges: e$_1'$ and e$_{\gamma+1}'$. Identifying outer
edges as explained above, leaves two outer edges open: e$_2$ and 
e$_{2\gamma-1}$. In this case we identify e$_1'$ with e$_{2\gamma-1}$ and 
e$_2$ with e$_{\gamma+1}'$, which again closes the invariant surface. Note that 
due to the identification of outer edges with inner edges, the central point of 
the rosette must be identified with the outermost points. The remaining points, 
must be identified as one single vertex $B$, if $\gamma$ is even [see 
figure~\ref{AP_fr}(b)], otherwise they constitute two vertices $B$ and $B'$ 
(this case is not shown). Hence the rosette (invariant surface) has $\gamma$ 
faces, $2\gamma$ edges, and $2$ vertices if $\gamma$ is even and $3$ vertices 
if $\gamma$ is odd. This gives $\chi= 2-\gamma \Rightarrow g= \gamma/2$ and 
$\chi=3-\gamma \Rightarrow g= (\gamma-1)/2$ respectively.

In all: $g=(\gamma-1)/2$ if $\gamma$ :odd, and $g=\gamma/2$ if $\gamma$
:even.  Hence $g= {\rm int}(\gamma/2)$.

\section*{References}

\bibliography{../../Bib/np,../../Bib/rm,../../Bib/sk,../../Bib/pp,../../Bib/lit,../../Bib/su}

\end{document}

%% file: deli12b.pstex_t
\begin{picture}(0,0)%
\epsfig{file=deli12b.pstex}%
\end{picture}%
\setlength{\unitlength}{1579sp}%
\begingroup\makeatletter\ifx\SetFigFont\undefined%
\gdef\SetFigFont#1#2#3#4#5{%
  \reset@font\fontsize{#1}{#2pt}%
  \fontfamily{#3}\fontseries{#4}\fontshape{#5}%
  \selectfont}%
\fi\endgroup%
\begin{picture}(9924,11912)(675,-11899)
\put(1126,-5464){\rotatebox{90.0}{\makebox(0,0)[b]{\smash{\SetFigFont{12}{14.4}{\rmdefault}{\mddefault}{\updefault}\special{ps: gsave 0 0 0 setrgbcolor}$\Delta I (s)$\special{ps: grestore}}}}}
\put(6382,-11899){\makebox(0,0)[b]{\smash{\SetFigFont{12}{14.4}{\rmdefault}{\mddefault}{\updefault}\special{ps: gsave 0 0 0 setrgbcolor}$s$\special{ps: grestore}}}}
\put(2089,-5462){\makebox(0,0)[rb]{\smash{\SetFigFont{6}{7.2}{\familydefault}{\mddefault}{\updefault}-0.1}}}
\put(2089,-4747){\makebox(0,0)[rb]{\smash{\SetFigFont{6}{7.2}{\familydefault}{\mddefault}{\updefault}-0.08}}}
\put(2089,-4032){\makebox(0,0)[rb]{\smash{\SetFigFont{6}{7.2}{\familydefault}{\mddefault}{\updefault}-0.06}}}
\put(2089,-3316){\makebox(0,0)[rb]{\smash{\SetFigFont{6}{7.2}{\familydefault}{\mddefault}{\updefault}-0.04}}}
\put(2089,-2601){\makebox(0,0)[rb]{\smash{\SetFigFont{6}{7.2}{\familydefault}{\mddefault}{\updefault}-0.02}}}
\put(2089,-1886){\makebox(0,0)[rb]{\smash{\SetFigFont{6}{7.2}{\familydefault}{\mddefault}{\updefault}0}}}
\put(2089,-1171){\makebox(0,0)[rb]{\smash{\SetFigFont{6}{7.2}{\familydefault}{\mddefault}{\updefault}0.02}}}
\put(2089,-456){\makebox(0,0)[rb]{\smash{\SetFigFont{6}{7.2}{\familydefault}{\mddefault}{\updefault}0.04}}}
\put(9376,-4636){\makebox(0,0)[rb]{\smash{\SetFigFont{6}{7.2}{\familydefault}{\mddefault}{\updefault}$3\pi/14$}}}
\put(9376,-4336){\makebox(0,0)[rb]{\smash{\SetFigFont{6}{7.2}{\familydefault}{\mddefault}{\updefault}$\pi/10$}}}
\put(9376,-5011){\makebox(0,0)[rb]{\smash{\SetFigFont{6}{7.2}{\familydefault}{\mddefault}{\updefault}GOE}}}
\put(9376,-3961){\makebox(0,0)[rb]{\smash{\SetFigFont{6}{7.2}{\familydefault}{\mddefault}{\updefault}irr}}}
\put(9376,-3661){\makebox(0,0)[rb]{\smash{\SetFigFont{6}{7.2}{\familydefault}{\mddefault}{\updefault}$3\pi/16$}}}
\put(9376,-3361){\makebox(0,0)[rb]{\smash{\SetFigFont{6}{7.2}{\familydefault}{\mddefault}{\updefault}$\pi/7$}}}
\put(9376,-3061){\makebox(0,0)[rb]{\smash{\SetFigFont{6}{7.2}{\familydefault}{\mddefault}{\updefault}$\pi/12$}}}
\put(9376,-2761){\makebox(0,0)[rb]{\smash{\SetFigFont{6}{7.2}{\familydefault}{\mddefault}{\updefault}$\pi/5$}}}
\put(9376,-2461){\makebox(0,0)[rb]{\smash{\SetFigFont{6}{7.2}{\familydefault}{\mddefault}{\updefault}$\pi/8$}}}
\put(2089,-11027){\makebox(0,0)[rb]{\smash{\SetFigFont{6}{7.2}{\familydefault}{\mddefault}{\updefault}-0.1}}}
\put(2089,-10312){\makebox(0,0)[rb]{\smash{\SetFigFont{6}{7.2}{\familydefault}{\mddefault}{\updefault}-0.08}}}
\put(2089,-9597){\makebox(0,0)[rb]{\smash{\SetFigFont{6}{7.2}{\familydefault}{\mddefault}{\updefault}-0.06}}}
\put(2089,-8881){\makebox(0,0)[rb]{\smash{\SetFigFont{6}{7.2}{\familydefault}{\mddefault}{\updefault}-0.04}}}
\put(2089,-8166){\makebox(0,0)[rb]{\smash{\SetFigFont{6}{7.2}{\familydefault}{\mddefault}{\updefault}-0.02}}}
\put(2089,-7451){\makebox(0,0)[rb]{\smash{\SetFigFont{6}{7.2}{\familydefault}{\mddefault}{\updefault}0}}}
\put(2089,-6736){\makebox(0,0)[rb]{\smash{\SetFigFont{6}{7.2}{\familydefault}{\mddefault}{\updefault}0.02}}}
\put(2089,-6021){\makebox(0,0)[rb]{\smash{\SetFigFont{6}{7.2}{\familydefault}{\mddefault}{\updefault}0.04}}}
\put(2208,-11226){\makebox(0,0)[b]{\smash{\SetFigFont{6}{7.2}{\familydefault}{\mddefault}{\updefault}0}}}
\put(3249,-11226){\makebox(0,0)[b]{\smash{\SetFigFont{6}{7.2}{\familydefault}{\mddefault}{\updefault}0.5}}}
\put(4291,-11226){\makebox(0,0)[b]{\smash{\SetFigFont{6}{7.2}{\familydefault}{\mddefault}{\updefault}1}}}
\put(5332,-11226){\makebox(0,0)[b]{\smash{\SetFigFont{6}{7.2}{\familydefault}{\mddefault}{\updefault}1.5}}}
\put(6373,-11226){\makebox(0,0)[b]{\smash{\SetFigFont{6}{7.2}{\familydefault}{\mddefault}{\updefault}2}}}
\put(7414,-11226){\makebox(0,0)[b]{\smash{\SetFigFont{6}{7.2}{\familydefault}{\mddefault}{\updefault}2.5}}}
\put(8456,-11226){\makebox(0,0)[b]{\smash{\SetFigFont{6}{7.2}{\familydefault}{\mddefault}{\updefault}3}}}
\put(9497,-11226){\makebox(0,0)[b]{\smash{\SetFigFont{6}{7.2}{\familydefault}{\mddefault}{\updefault}3.5}}}
\put(10538,-11226){\makebox(0,0)[b]{\smash{\SetFigFont{6}{7.2}{\familydefault}{\mddefault}{\updefault}4}}}
\end{picture}

%% file: kmatmap.pstex_t
\begin{picture}(0,0)%
\epsfig{file=kmatmap.pstex}%
\end{picture}%
\setlength{\unitlength}{1579sp}%
\begingroup\makeatletter\ifx\SetFigFont\undefined%
\gdef\SetFigFont#1#2#3#4#5{%
  \reset@font\fontsize{#1}{#2pt}%
  \fontfamily{#3}\fontseries{#4}\fontshape{#5}%
  \selectfont}%
\fi\endgroup%
\begin{picture}(12109,5424)(1104,-5773)
\put(3211,-668){\makebox(0,0)[lb]{\smash{\SetFigFont{10}{12.0}{\rmdefault}{\mddefault}{\updefault}$\vec y_1$}}}
\put(12961,-661){\makebox(0,0)[lb]{\smash{\SetFigFont{10}{12.0}{\rmdefault}{\mddefault}{\updefault}$\vec y_0$}}}
\put(2551,-2011){\makebox(0,0)[rb]{\smash{\SetFigFont{10}{12.0}{\rmdefault}{\mddefault}{\updefault}$a_0$}}}
\put(2551,-3661){\makebox(0,0)[rb]{\smash{\SetFigFont{10}{12.0}{\rmdefault}{\mddefault}{\updefault}$a_1$}}}
\put(2551,-4636){\makebox(0,0)[rb]{\smash{\SetFigFont{10}{12.0}{\rmdefault}{\mddefault}{\updefault}$a_{-1}$}}}
\end{picture}

%% file: mapflow.pstex_t
\begin{picture}(0,0)%
\epsfig{file=mapflow.pstex}%
\end{picture}%
\setlength{\unitlength}{1105sp}%
\begingroup\makeatletter\ifx\SetFigFont\undefined%
\gdef\SetFigFont#1#2#3#4#5{%
  \reset@font\fontsize{#1}{#2pt}%
  \fontfamily{#3}\fontseries{#4}\fontshape{#5}%
  \selectfont}%
\fi\endgroup%
\begin{picture}(12958,1368)(639,-2395)
\put(7651,-1711){\makebox(0,0)[b]{\smash{\SetFigFont{8}{9.6}{\rmdefault}{\mddefault}{\updefault}
$a_0$
}}}
\put(11251,-1711){\makebox(0,0)[b]{\smash{\SetFigFont{8}{9.6}{\rmdefault}{\mddefault}{\updefault}
$a_2$
}}}
\put(13051,-1711){\makebox(0,0)[b]{\smash{\SetFigFont{8}{9.6}{\rmdefault}{\mddefault}{\updefault}
$a_3$
}}}
\put(5701,-1711){\makebox(0,0)[b]{\smash{\SetFigFont{8}{9.6}{\rmdefault}{\mddefault}{\updefault}
$a_{-1}$
}}}
\put(3601,-1711){\makebox(0,0)[b]{\smash{\SetFigFont{8}{9.6}{\rmdefault}{\mddefault}{\updefault}
$a_{-2}$
}}}
\put(1501,-1711){\makebox(0,0)[b]{\smash{\SetFigFont{8}{9.6}{\rmdefault}{\mddefault}{\updefault}
$a_{-3}$
}}}
\put(9451,-1711){\makebox(0,0)[b]{\smash{\SetFigFont{8}{9.6}{\rmdefault}{\mddefault}{\updefault}
$a_1$
}}}
\end{picture}

%% file: CM.pstex_t
\begin{picture}(0,0)%
\epsfig{file=CM.pstex}%
\end{picture}%
\setlength{\unitlength}{1579sp}%
\begingroup\makeatletter\ifx\SetFigFont\undefined%
\gdef\SetFigFont#1#2#3#4#5{%
  \reset@font\fontsize{#1}{#2pt}%
  \fontfamily{#3}\fontseries{#4}\fontshape{#5}%
  \selectfont}%
\fi\endgroup%
\begin{picture}(9964,12158)(632,-12091)
\put(1083,-5466){\rotatebox{90.0}{\makebox(0,0)[b]{\smash{\SetFigFont{12}{14.4}{\rmdefault}{\mddefault}{\updefault}\special{ps: gsave 0 0 0 setrgbcolor}$\Delta I (s)$\special{ps: grestore}}}}}
\put(6371,-12091){\makebox(0,0)[b]{\smash{\SetFigFont{12}{14.4}{\rmdefault}{\mddefault}{\updefault}\special{ps: gsave 0 0 0 setrgbcolor}$s$\special{ps: grestore}}}}
\put(9495,-6217){\makebox(0,0)[lb]{\smash{\SetFigFont{12}{14.4}{\rmdefault}{\mddefault}{\updefault}\special{ps: gsave 0 0 0 setrgbcolor}(b)\special{ps: grestore}}}}
\put(9498,-631){\makebox(0,0)[lb]{\smash{\SetFigFont{12}{14.4}{\rmdefault}{\mddefault}{\updefault}\special{ps: gsave 0 0 0 setrgbcolor}(a)\special{ps: grestore}}}}
\put(2086,-11035){\makebox(0,0)[rb]{\smash{\SetFigFont{6}{7.2}{\familydefault}{\mddefault}{\updefault}-0.1}}}
\put(2086,-9962){\makebox(0,0)[rb]{\smash{\SetFigFont{6}{7.2}{\familydefault}{\mddefault}{\updefault}-0.05}}}
\put(2086,-8889){\makebox(0,0)[rb]{\smash{\SetFigFont{6}{7.2}{\familydefault}{\mddefault}{\updefault}0}}}
\put(2086,-7817){\makebox(0,0)[rb]{\smash{\SetFigFont{6}{7.2}{\familydefault}{\mddefault}{\updefault}0.05}}}
\put(2086,-6744){\makebox(0,0)[rb]{\smash{\SetFigFont{6}{7.2}{\familydefault}{\mddefault}{\updefault}0.1}}}
\put(2086,-5671){\makebox(0,0)[rb]{\smash{\SetFigFont{6}{7.2}{\familydefault}{\mddefault}{\updefault}0.15}}}
\put(2205,-11234){\makebox(0,0)[b]{\smash{\SetFigFont{6}{7.2}{\familydefault}{\mddefault}{\updefault}0}}}
\put(3246,-11234){\makebox(0,0)[b]{\smash{\SetFigFont{6}{7.2}{\familydefault}{\mddefault}{\updefault}0.5}}}
\put(4288,-11234){\makebox(0,0)[b]{\smash{\SetFigFont{6}{7.2}{\familydefault}{\mddefault}{\updefault}1}}}
\put(5329,-11234){\makebox(0,0)[b]{\smash{\SetFigFont{6}{7.2}{\familydefault}{\mddefault}{\updefault}1.5}}}
\put(6370,-11234){\makebox(0,0)[b]{\smash{\SetFigFont{6}{7.2}{\familydefault}{\mddefault}{\updefault}2}}}
\put(7411,-11234){\makebox(0,0)[b]{\smash{\SetFigFont{6}{7.2}{\familydefault}{\mddefault}{\updefault}2.5}}}
\put(8453,-11234){\makebox(0,0)[b]{\smash{\SetFigFont{6}{7.2}{\familydefault}{\mddefault}{\updefault}3}}}
\put(9494,-11234){\makebox(0,0)[b]{\smash{\SetFigFont{6}{7.2}{\familydefault}{\mddefault}{\updefault}3.5}}}
\put(10535,-11234){\makebox(0,0)[b]{\smash{\SetFigFont{6}{7.2}{\familydefault}{\mddefault}{\updefault}4}}}
\put(2089,-4389){\makebox(0,0)[rb]{\smash{\SetFigFont{6}{7.2}{\familydefault}{\mddefault}{\updefault}-0.05}}}
\put(2089,-3316){\makebox(0,0)[rb]{\smash{\SetFigFont{6}{7.2}{\familydefault}{\mddefault}{\updefault}0}}}
\put(2089,-2244){\makebox(0,0)[rb]{\smash{\SetFigFont{6}{7.2}{\familydefault}{\mddefault}{\updefault}0.05}}}
\put(2089,-1171){\makebox(0,0)[rb]{\smash{\SetFigFont{6}{7.2}{\familydefault}{\mddefault}{\updefault}0.1}}}
\put(2089,-98){\makebox(0,0)[rb]{\smash{\SetFigFont{6}{7.2}{\familydefault}{\mddefault}{\updefault}0.15}}}
\end{picture}

%% file: f14fin.pstex_t
\begin{picture}(0,0)%
\epsfig{file=f14fin.pstex}%
\end{picture}%
\setlength{\unitlength}{1579sp}%
\begingroup\makeatletter\ifx\SetFigFont\undefined%
\gdef\SetFigFont#1#2#3#4#5{%
  \reset@font\fontsize{#1}{#2pt}%
  \fontfamily{#3}\fontseries{#4}\fontshape{#5}%
  \selectfont}%
\fi\endgroup%
\begin{picture}(9594,7260)(1199,-7579)
\put(9214,-7177){\makebox(0,0)[b]{\smash{\SetFigFont{6}{7.2}{\rmdefault}{\mddefault}{\updefault}\special{ps: gsave 0 0 0 setrgbcolor}0\special{ps: grestore}}}}
\put(7699,-7177){\makebox(0,0)[b]{\smash{\SetFigFont{6}{7.2}{\rmdefault}{\mddefault}{\updefault}\special{ps: gsave 0 0 0 setrgbcolor}-2\special{ps: grestore}}}}
\put(8457,-7177){\makebox(0,0)[b]{\smash{\SetFigFont{6}{7.2}{\rmdefault}{\mddefault}{\updefault}\special{ps: gsave 0 0 0 setrgbcolor}-1\special{ps: grestore}}}}
\put(9970,-7174){\makebox(0,0)[b]{\smash{\SetFigFont{6}{7.2}{\rmdefault}{\mddefault}{\updefault}\special{ps: gsave 0 0 0 setrgbcolor}1\special{ps: grestore}}}}
\put(10732,-7177){\makebox(0,0)[b]{\smash{\SetFigFont{6}{7.2}{\rmdefault}{\mddefault}{\updefault}\special{ps: gsave 0 0 0 setrgbcolor}2\special{ps: grestore}}}}
\put(9214,-7579){\makebox(0,0)[b]{\smash{\SetFigFont{6}{7.2}{\rmdefault}{\mddefault}{\updefault}\special{ps: gsave 0 0 0 setrgbcolor}$\times 10^{-4}$\special{ps: grestore}}}}
\put(6450,-1315){\makebox(0,0)[lb]{\smash{\SetFigFont{6}{7.2}{\familydefault}{\mddefault}{\updefault}5}}}
\put(6450,-2279){\makebox(0,0)[lb]{\smash{\SetFigFont{6}{7.2}{\familydefault}{\mddefault}{\updefault}10}}}
\put(6450,-3273){\makebox(0,0)[lb]{\smash{\SetFigFont{6}{7.2}{\familydefault}{\mddefault}{\updefault}15}}}
\put(6450,-4223){\makebox(0,0)[lb]{\smash{\SetFigFont{6}{7.2}{\familydefault}{\mddefault}{\updefault}20}}}
\put(6450,-5157){\makebox(0,0)[lb]{\smash{\SetFigFont{6}{7.2}{\familydefault}{\mddefault}{\updefault}25}}}
\put(6450,-6121){\makebox(0,0)[lb]{\smash{\SetFigFont{6}{7.2}{\familydefault}{\mddefault}{\updefault}30}}}
\put(1261,-7198){\makebox(0,0)[b]{\smash{\SetFigFont{6}{7.2}{\familydefault}{\mddefault}{\updefault}0}}}
\put(2110,-7198){\makebox(0,0)[b]{\smash{\SetFigFont{6}{7.2}{\familydefault}{\mddefault}{\updefault}50}}}
\put(2959,-7198){\makebox(0,0)[b]{\smash{\SetFigFont{6}{7.2}{\familydefault}{\mddefault}{\updefault}100}}}
\put(3808,-7198){\makebox(0,0)[b]{\smash{\SetFigFont{6}{7.2}{\familydefault}{\mddefault}{\updefault}150}}}
\put(4656,-7198){\makebox(0,0)[b]{\smash{\SetFigFont{6}{7.2}{\familydefault}{\mddefault}{\updefault}200}}}
\put(5505,-7198){\makebox(0,0)[b]{\smash{\SetFigFont{6}{7.2}{\familydefault}{\mddefault}{\updefault}250}}}
\put(6450,-545){\makebox(0,0)[lb]{\smash{\SetFigFont{6}{7.2}{\familydefault}{\mddefault}{\updefault}1}}}
\end{picture}